\documentclass[twocolumn,preprintnumbers,amsmath,amssymb]{revtex4-1}
\usepackage{graphicx}% Include figure files
\usepackage{dcolumn}% Align table columns on decimal point
\usepackage{bm}% bold math
\usepackage{textcomp}
\usepackage{amssymb}
\usepackage{color}

\begin{document}

\title{Quantum-Limited Amplification of Cavity Optomechanics without Resolved Sideband Condition}
\author{Wen-Juan Yang$ ^{1,2}$, Yu-Kai Wu$ ^{1}$}, \author{Xiang-Bin Wang$ ^{1,2,3}$}\email{Email
Address:xbwang@mail.tsinghua.edu.cn}
\affiliation{ \centerline{$^{1}$State Key Laboratory of Low
Dimensional Quantum Physics, Tsinghua University, Beijing 100084,
People's Republic of China}\centerline{$^{2}$Synergetic Innovation Center of Quantum Information and Quantum Physics,}\centerline{University of Science and Technology of China, Hefei, Anhui 230026, China}\centerline{$^{3}$Shandong
Academy of Information and Communication Technology, Jinan 250101,
People's Republic of China}}

\date{\today}
\begin{abstract}
 We propose a scheme to realize the phase-preserving amplification without the restriction of resolved sideband condition. As a result, our gain-bandwidth product is about one magnitude larger than the existing proposals. In our model, an additional cavity is coupled to the cavity-optomechanical system. Therefore our operating frequency is continuously tunable via adjusting the coupling coefficient of the two cavities.
\end{abstract}
\maketitle
\parskip=0 pt
\section{Introduction}
In the past few years, significant progress has been achieved in the cavity optomechanics\cite{co1, co2, co3, co4}. Examples include ground state cooling of a mechanical oscillator which is a prerequisite for its applications in quantum information processing\cite{cooling1, cooling2,cooling3,cooling4,cooling5,cooling6} and sensitive measurement\cite{sem1, am1, am2, sem2}, and its serving as intermediate transducer of hybrid systems\cite{sem2, hs1, hs2, hs3}.

%The sensitive measurement is an important issue in modern physics\cite{mp}. Due to the zero-point fluctuations, the noise added by a linear amplifier equals at least the quantum fluctuation of the system}\cite{qf}.
To reduce noise in sensitive measurement, there have been great efforts to reach the quantum limit of phase-preserving amplifier, which arise from the zero-point fluctuation\cite{mp, qf}. Substantial progress has been achieved in superconducting systems with Josephson ring\cite{sc1, sc2}. In cavity optomechanical systems, there are also schemes discussing phase-preserving amplifier, all of which however, work in the resolved sideband regime\cite{am1, am2}.
%There has} been substantial progress in realizing quantum-limited and phase-preserving amplifier in superconducting devices with Josephson ring\cite{sc1, sc2}. There are also} schemes discussing phase-preserving amplifier in cavity optomechanics systems\cite{am1, am2}, all of which however,} work in the resolved sideband regime of cavity optomechanics systems.
In practice, it is difficult to reach the resolved sideband regime, say $\kappa\ll\omega_m$ where $\kappa$ is the dissipation rate of the cavity and $\omega_m$ is the frequency of the mechanical oscillator\cite{us1, us2}. Here we propose to realize the phase-preserving amplification beyond the resolved sideband condition. In our model, an auxiliary cavity is coupled to the cavity-optomechanical system. Since we work in a much larger parameter regime, the optimal gain-bandwidth product is enhanced by one magnitude. By means of adjusting the two-cavity coupling coefficient\cite{tu1, tu2, tu3, tu4, tu5, tu6, tu7, tu8}, the operating frequency in our scheme is tunable. This is an important advantage over previous proposals where the operating frequency is not adjustable.

The paper is organized as follows. Section \uppercase \expandafter{\romannumeral 2} presents the model, its Hamiltonian, and the analytical formulas of the amplification process. In Section \uppercase \expandafter{\romannumeral 3} we optimize the parameters within and beyond the resolved-sideband regime numerically. Then we discuss the adjustability of the amplifier's center frequency. We give the conclusion in Sec. \uppercase \expandafter{\romannumeral 4}.

\section{model and formulas}

\begin{figure}[htbp]
  \centering
  \includegraphics[width=0.4\textwidth]{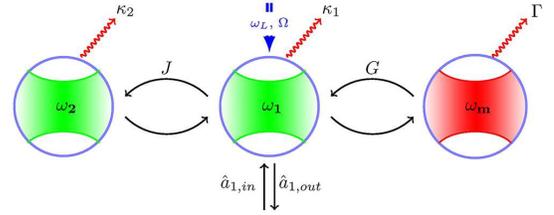}\\
  \caption{The schematic of the cavity optomechanical system coupled with an auxiliary cavity.}\label{model}
\end{figure}

A schematic of our model is shown in Fig. \ref{model}. An auxiliary cavity is coupled to a cavity-optomechanical system.  The Hamiltonian of the compound cavity system is
\begin{eqnarray}
H&=&\omega_1 \hat{a}_1^{\dagger}\hat{a}_1+\omega_2 \hat{a}_2^{\dagger}\hat{a}_2+J(\hat{a}_1^{\dagger}\hat{a}_2+\hat{a}_2^\dagger \hat{a}_1)+\omega_m \hat{b}^{\dagger}\hat{b} \notag\\
&&-g_0(\hat{b}^{\dagger}+\hat{b})\hat{a}_1^\dagger \hat{a}_1+\Omega(\hat{a}_1^\dagger e^{-i\omega_L t}+\hat{a}_1 e^{i\omega_L t}),
\end{eqnarray}
where $\hat{a}_1$ and $\hat{a}_2$ are the annihilation operators of the two cavity modes with $\omega_1$ and $\omega_2$ being their frequencies, $J$ is the coupling coefficient of the two cavity modes, $b$ is the annihilation operator of the mechanical oscillator with frequency $\omega_m$, $g_0$ is the parametric coupling strength between the first cavity and the mechanical oscillator, $\Omega$ is the strength of the external driving field for the first cavity, and $\omega_L$ is the frequency of the driving field.

Moving into the frame of the drive, the Hamiltonian reads
\begin{eqnarray}
\tilde{H}&=&-\Delta_{10} \hat{a}_1^{\dagger}\hat{a}_1-\Delta_2 \hat{a}_2^{\dagger}\hat{a}_2+J(\hat{a}_1^{\dagger}\hat{a}_2+\hat{a}_2^\dagger \hat{a}_1) \notag\\
&&+\omega_m \hat{b}^{\dagger}\hat{b}-g_0(\hat{b}^{\dagger}+\hat{b})\hat{a}_1^\dagger \hat{a}_1+\Omega(\hat{a}_1^\dagger + \hat{a}_1),
\end{eqnarray}
where $\Delta_{10}=\omega_L-\omega_1$ and $\Delta_2=\omega_L-\omega_2$ are the detunings between the two cavity modes and the driving field.

We can linearize the dynamical equations of the driven compound cavity system by assuming $\hat{a}_1=\bar{a}_1+\delta\hat{a}_1$, $\hat{a}_2=\bar{a}_2+\delta\hat{a}_2$ and $\hat{b}=\bar{b}+\delta\hat{b}$ where $\bar{a}_1$, $\bar{a}_2$ and $\bar{b}$ are the respective mean values.
%obey the following nonlinear classical equations of motion,
%\begin{eqnarray}
%\dot{\bar{a}}_1&=&i\Delta_{10}\bar{a}_1}-\frac{\kappa_1}{2}\bar{a}_1-iJ\bar{a}_2+ig_0(\bar{b}+\bar{b}^{*})\bar{a}_1-i\Omega, \\
%\dot{\bar{a}}_2&=&i\Delta_2\bar{a}_2}-\frac{\kappa_2}{2}\bar{a}_2-iJ\bar{a}_1, \\
%\dot{\bar{b}}&=&-i\omega_m\bar{b}-\frac{\Gamma}{2}\bar{b}+ig_0|\bar{a}_1|^2,
%\end{eqnarray}
Neglecting nonlinear terms we get
\begin{eqnarray}
\dot{\delta\hat{a}}_1&=&i\Delta_1\delta\hat{a}_1-\frac{\kappa_1}{2}\delta\hat{a}_1-iJ\delta\hat{a}_2+iG(\delta\hat{b}+\delta\hat{b}^\dagger)\notag\\
&&+\sqrt{\kappa_1}\hat{a}_{1,in}, \label{a1}\\
\dot{\delta\hat{a}}_2&=&i\Delta_2\delta\hat{a}_2-\frac{\kappa_2}{2}\delta\hat{a}_2-iJ\delta\hat{a}_1+\sqrt{\kappa_2}\hat{a}_{2,in}, \label{a2}\\
\dot{\delta\hat{b}}&=&-i\omega_m\delta\hat{b}-\frac{\Gamma}{2}\delta\hat{b}+iG(\delta\hat{a}_1+\delta\hat{a}_1^\dagger)+\sqrt{\Gamma}\hat{b}_{in} \label{b},
\end{eqnarray}
where $\Delta_1=\Delta_{10}+g_0(\bar{b}+\bar{b}^{*})$ denotes the effective detuning of the first cavity, $G=g_0\bar{a}_1$ is the enhanced optomechanical coupling, $\kappa_1$, $\kappa_2$ and $\Gamma$ are the decay rates of cavity $1$, $2$ and the mechanical oscillator.

We write Eqs. (\ref{a1}), (\ref{a2}) and (\ref{b}) in the matrix form
\begin{equation}
\dot{\textbf{u}}=\textbf{M}\,\textbf{u}+\textbf{L}\,\textbf{u}_{\textbf{in}},
\end{equation}
where
\begin{eqnarray*}
\textbf{u}&=&(\delta\hat{a}_1, \delta\hat{a}_1^\dagger, \delta\hat{a}_2, \delta\hat{a}_2^\dagger, \delta\hat{b}, \delta\hat{b}^\dagger)^{\emph{T}},\\
\textbf{u}_{\textbf{in}}&=&(\hat{a}_{1,in}, \hat{a}_{1,in}^\dagger, \hat{a}_{2,in}, \hat{a}_{2,in}^\dagger, \hat{b}_{in}, \hat{b}_{in}^\dagger)^{\emph{T}},
\end{eqnarray*}
\begin{widetext}
\begin{equation}
\textbf{M}=\left(
\begin{array}{cccccc}
i\Delta_1-\frac{\kappa_1}{2} & 0 & -iJ & 0 & iG & iG\\
0 & -i\Delta_1-\frac{\kappa_1}{2} & 0 & iJ & -iG & -iG\\
-iJ & 0 & i\Delta_2-\frac{\kappa_2}{2} & 0 & 0 & 0\\
0 & iJ & 0 & -i\Delta_2-\frac{\kappa_2}{2} & 0 & 0\\
iG & iG & 0 & 0 & -i\omega_m-\frac{\Gamma}{2} & 0\\
-iG & -iG & 0 &0 & 0 &i\omega_m -\frac{\Gamma}{2}\end{array}
\right),
\end{equation}
\end{widetext}
and $\textbf{L}=\texttt{Diag}[\sqrt{\kappa_1}, \sqrt{\kappa_1}, \sqrt{\kappa_2}, \sqrt{\kappa_2}, \sqrt{\Gamma}, \sqrt{\Gamma}]$.

We work in frequency space and use the input-output relations $\textbf{u}_{\textbf{out}}=\textbf{u}_{\textbf{in}}-\textbf{L}\,\textbf{u}$. Then we get
\begin{equation}
\textbf{u}_{\textbf{out}}[\omega]=\textbf{U}(\omega)\,\textbf{u}_{\textbf{in}}[\omega]
\end{equation}
with
\begin{equation}
\textbf{U}=\textbf{1}+\textbf{L}\,[i\omega+\textbf{M}]^{-1}\,\textbf{L}.
\end{equation}
The amplification can be expressed as
\begin{eqnarray}
\hat{a}_{1,out}&=&A(\omega)\hat{a}_{1,in}+B(\omega)\hat{a}_{1,in}^{\dagger}+C(\omega)\hat{a}_{2,in}\notag\\
&&+D(\omega)\hat{a}_{2,in}^\dagger+E(\omega)\hat{b}_{in}+F(\omega)\hat{b}_{in}^\dagger.
\end{eqnarray}
In the case of unresolved sideband for the first cavity where $\kappa_1\geq\omega_m$, we cannot make the resolved sideband approximation. To simplify, we first consider the case where $\kappa_2=0$. The power gain $\mathcal{G}[\omega]=|A(\omega)|^2$ can be expressed as
\begin{widetext}
\begin{eqnarray}
\mathcal{G}[\omega]=\left|1+\frac{2i\kappa_1(-\Delta_2-\omega)\{[-2J^2+(-\Delta_2+\omega)(-2\Delta_1+i\kappa_1+2\omega)]\alpha(\omega)
-16G^2\omega_m(-\Delta_2+\omega)\}}{\alpha(\omega)\{4J^4+(\Delta_2^2-\omega^2)[(\kappa_1-2i\omega)^2+4\Delta_1^2]-4iJ^2\omega(\kappa_1-2i\omega)
-8\Delta_1\Delta_2J^2\}+64G^2\omega_m\beta(\omega)}\right|^2,
\label{g}
\end{eqnarray}
\end{widetext}
where $\alpha(\omega)=4\omega_m^2+(\Gamma-2i\omega)^2$ and $\beta(\omega)=\Delta_1(\Delta_2^2-\omega^2)-\Delta_2J^2$. We discuss on the blue sideband where $\Delta_1=\omega_m$. To suppress the amplification in the vicinity of $\omega_m$, we simply choose $\Delta_2=-\omega_m$. Denote the denominator of the second term in the right hand side of Eq. (\ref{g}) as $\rho(\omega)$. The critical points of instability appear at $\rho(\omega)=0$.
%which is equivalent to the Routh-Hurwitz criterion. This criterion is used to evaluate the stability of amplification process\cite{rhc}.

\begin{figure}[!hbp]
  \includegraphics[width=0.45\textwidth]{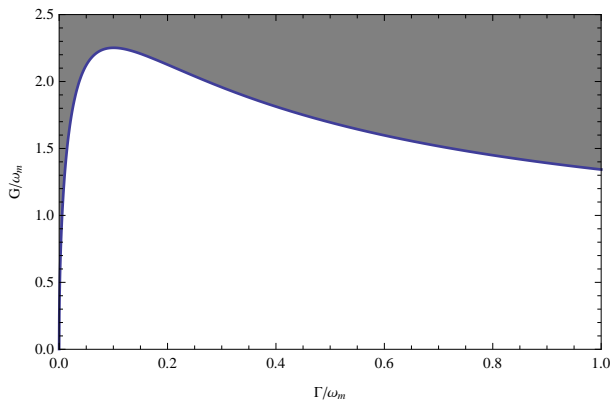}
  \caption{The gray area is the unstable regime. Here $J=3\omega_m$ and $\kappa_1=0.1\omega_m$.}\label{unstable}
\end{figure}

According to this, we get two groups of boundary conditions:
\begin{eqnarray}
G^{(1)}&=&\frac{\sqrt{\Gamma\kappa _1}\sqrt{(4J^2+\Gamma^2+\Gamma\kappa_1)^2+16\omega _m^2(\Gamma+\kappa_1)^2}}{8(\Gamma+\kappa_1)\omega_m},\notag\\
\omega^{(1)}&=&\pm\sqrt{J^2\Gamma\left/\left(\Gamma +\kappa _1\right)\right.+\omega_m^2};\label{omega1}\\
G^{(2)}&=&\frac{\sqrt{\kappa _1}\sqrt{(4J^2+\Gamma^2+\Gamma\kappa_1)^2+16\Gamma^2\omega_m^2}}{8\sqrt{\Gamma}\omega_m},\notag\\
\omega^{(2)}&=&\pm\sqrt{J^2+\Gamma\kappa _1/4+\omega_m^2}.
\end{eqnarray}
We plot the stable and unstable boundary $G^{(1)}(\Gamma)$ in Fig. \ref{unstable}.
The gray area is the unstable regime. Applying the Routh-Hurwitz criterion\cite{rhc}, $G^{(2)}(\Gamma)$ is the separation of two-negative roots and four negative roots. In both regimes the system is unstable. So we do not draw it here.

Consider a set of parameters in the stable regime where $G$ deviates by a small quantity $-\delta$($\delta>0$) from $G^{(1)}$, which is calculated from other parameters. We can always write $\mathcal{G}[\omega]$ in the vicinity of $-|\omega^{(1)}|$ as
\begin{eqnarray}
\mathcal{G}[\omega]=\left|1+\frac{\kappa}{x+a}\right|^2,
\end{eqnarray}
where $x$ is the deviation of $\omega$ from $-|\omega^{(1)}|$. We can easily get that
\begin{eqnarray}
\kappa &=& \frac{\mathrm{i}\kappa_1 J^2\Gamma (\xi - 4 \mathrm{i}\omega_m)(\Gamma \xi + 8\omega_m^2 + 8\omega_m\sigma)}{4\omega_m\xi \sigma \left[ \left(\Gamma + \kappa_1\right)^2\left(4\mathrm{i}\sigma + \Gamma\right) - 4 J^2 \left(\Gamma - \kappa_1\right) \right]},\\
a &=& \frac{-8 J^2 \omega_m\sqrt{\Gamma} \sqrt{\kappa_1} \sqrt{\xi^2 + 16\omega_m^2} \delta}{\xi \sigma \left[\left(\Gamma + \kappa_1 \right)^2 \left(4\mathrm{i} \sigma + \Gamma\right) - 4 J^2 \left(\Gamma - \kappa_1\right) \right]}\label{omegadv}.
\end{eqnarray}
where $\xi = \Gamma + 4J^2 / (\Gamma+\kappa_1)$, $\sigma = \sqrt{\omega_m^2+J^2 \Gamma/(\Gamma+\kappa_1)}$. Here -$\mathrm{Re}[a]$ is the deviation of the peak from $-|\omega^{(1)}|$. The bandwidth of the amplifier is given by $2\mathrm{Im}[a]$ and the gain-bandwidth product is $|\kappa|$.

\section{discussion}
\subsection{resolved sideband regime}
\begin{figure}[htbp]
  \centering
  \includegraphics[width=0.5\textwidth]{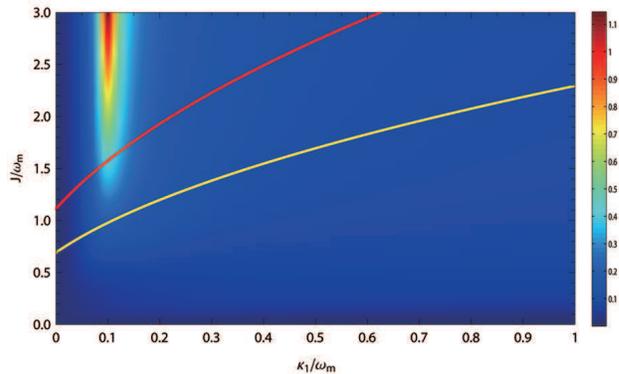}
  \caption{Gain-bandwidth product $|\kappa|/\omega_m$ in the $J-\kappa_1$ plane. The yellow line and red line are $|B|/|A|=0.1$ and $0.2$ respectively. Here $\Gamma=0.1\omega_m$.}\label{optimizedkappa}
\end{figure}

\begin{figure}[htbp]
  \includegraphics[width=0.4\textwidth]{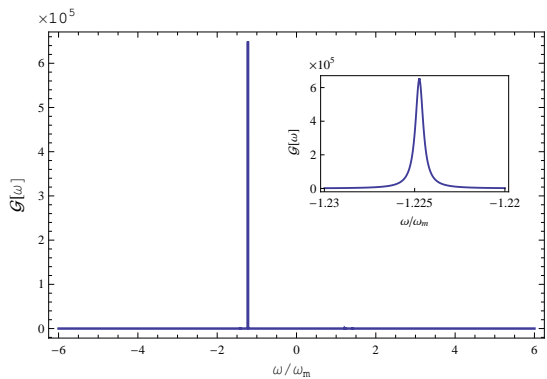}
  \caption{Power gain $\mathcal{G}[\omega]$ as a function of the frequency $\omega$ for $J=1.0\omega_m$, $\kappa_1=0.1\omega_m$, $\kappa_2=0$, $G=0.2561\omega_m$ and $\Gamma_m=0.1\omega_m$.}\label{kappa20}
\end{figure}

\begin{figure}[htbp]
  \includegraphics[width=0.58\textwidth]{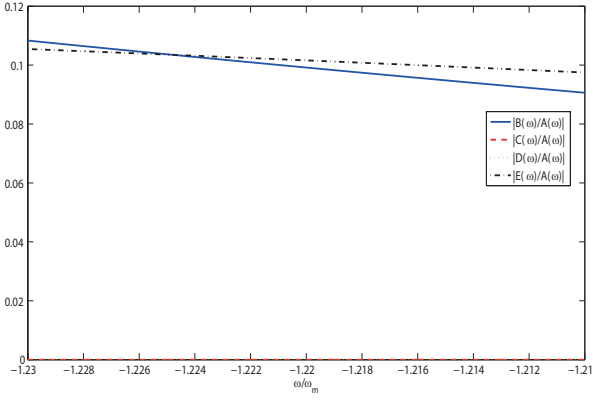}
  \caption{$|B(\omega|/|A(\omega)|$(blue solid line), $|C(\omega)|/|A(\omega)|$(red dashed line), $|D(\omega)|/|A(\omega)|$(green dotted line) and $|E(\omega)|/|A(\omega)|$(black dash dot line) as functions of the frequency $\omega$. The parameters are the same as in Fig. \ref{kappa20}.}\label{kappa2020}
\end{figure}

\begin{figure}[htbp]
  \includegraphics[width=0.68\textwidth]{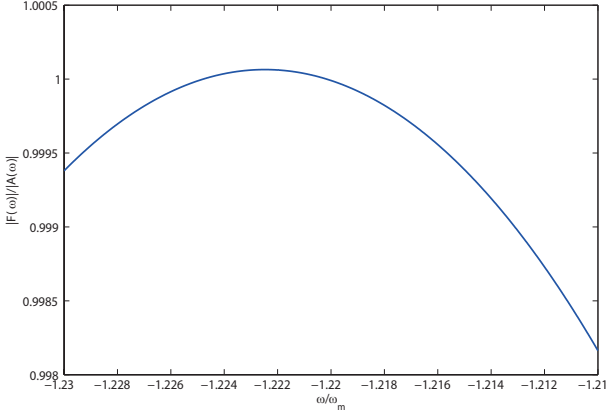}\\
  \caption{$|F(\omega)|/|A(\omega)|$ as a function of the frequency $\omega$. The parameters are the same as in Fig. \ref{kappa20}. }\label{kappa20contrast}
\end{figure}

In Fig. \ref{optimizedkappa} we plot the gain-bandwidth product $|\kappa|/\omega_m$ in the $J-\kappa_1$ plane where $\Gamma=0.1\omega_m$(colored image). The optimal $|\kappa|$ is found at $\kappa_1=\Gamma$. $|B|/|A|=0.1$(yellow line) and 0.2(red line) are shown in this figure. We choose $J=1.0\omega_m$, $\kappa_1=0.1\omega_m$, $\Gamma=0.1\omega_m$ and $G=G^{(1)}-0.0001\omega_m=0.2561\omega_m$. In this case the maximum power gain appears at $\omega_{p}=-1.2247\omega_m$, the bandwidth is $5.2\times10^{-4}\omega_m$ and the gain-bandwidth product is $|\kappa|=0.21\omega_m$. We plot the power gain $\mathcal{G}[\omega]$ as a function of the frequency $\omega$ in Fig. \ref{kappa20}.
In Fig. \ref{kappa2020} we plot $|B(\omega)|/|A(\omega)|, |C(\omega)|/|A(\omega)|, |D(\omega)|/|A(\omega)|$ and $|E(\omega)|/|A(\omega)|$ as functions
of the frequency $\omega$ in the vicinity of the peak.
where we can make an approximation that $|B(\omega)|, |C(\omega)|, |D(\omega)|, |E(\omega)|\ll|A(\omega)|$. As
a consequence, the amplifier is phase insensitive\cite{mp, ps}. We plot $|F(\omega)|/|A(\omega)|$ as a function of the frequency $\omega$ in Fig. \ref{kappa20contrast}. It can be proved that at the peak of amplification, we have $|F|=|A|$. According to this, the number of added noise photons
\begin{equation}
N(\omega_p)=\frac{|F|^2}{|A|^2}(\bar{n}_{eff}+\frac{1}{2})\approx \bar{n}_{eff}+\frac{1}{2},
\end{equation}
where $\bar{n}_{eff}$ is the effective phonon number of the reservoir of the oscillator. We can use the method guided by Ref. \cite{am1} to cool the mechanical resonator to the quantum ground state. As $|F|^2/|A|^2\approx1$ and $\bar{n}_{eff}\rightarrow 0$, the amplifier is quantum limited.

Now we take the non-zero dissipation of cavity 2($\kappa_2\neq 0$) into consideration where $\kappa_2=0.01\omega_m$, $J=1.0\omega_m$, $\kappa_1=0.1\omega_m$, $\Gamma=0.1\omega_m$ and $G=0.2561\omega_m$. We can also get a quantum-limited and phase-preserving amplifier. the maximum gain power appears at $\omega_{p}=-1.22\omega_m$, the bandwidth is $2.1\times10^{-3}\omega_m$ and the gain-bandwidth product $|\kappa|=0.21\omega_m$. The optimal gain-bandwidth product is $|\kappa|=0.43\omega_m$ if we choose $|B|/|A|=0.2$. In previous work Ref. \cite{am1}, the gain-bandwidth product is given by the cavity linwidth $\kappa$ and is restricted by the resolved sideband condition $\kappa \ll \omega_m$. Therefore our result is at least one magnitude better.
\subsection{beyond resolved sideband condition}
\begin{figure}[htbp]
  \centering
  \includegraphics[width=0.4\textwidth]{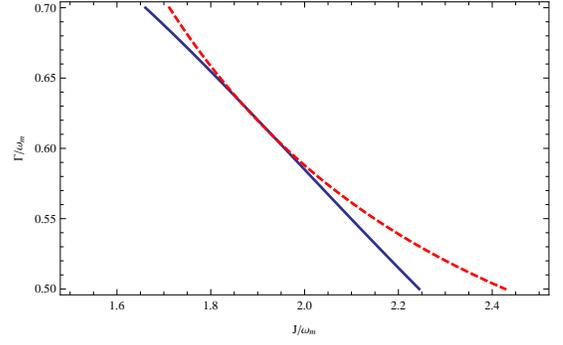}\\
  \caption{Contours of $|B|/|A|=0.2$(blue solid line) and $|\kappa|=0.5\omega_m$(red dashed line). Here $\kappa_1=2\omega_m$, $\kappa_2=0$ and $G=G^{1}-0.0001\omega_m$.}\label{unresolved}
\end{figure}

In this subsection we discard the resolved sideband condition and further optimize the parameters. For discussion simplicity, here we fix $\kappa_1=2.0\omega_m$. In Fig. \ref{unresolved} we plot the contours of $|B|/|A|=0.2$(blue solid line) and $|\kappa|=0.5\omega_m$(red dashed line) in the  $J-\Gamma$ plane.
 We roughly choose $\kappa_1=2.0\omega_m$, $\kappa_2=0$, $J=2.0\omega_m$, $\Gamma=0.6\omega_m$ and $G=G^{(1)}-0.0001\omega_m=1.0747\omega_m$. The maximum gain power appears at $\omega_{p}=-1.39\omega_m$, the bandwidth is $1.0\times10^{-4}\omega_m$ and the gain-bandwidth product is $|\kappa|=0.51\omega_m$.

\begin{figure}
  \begin{flushleft}
  \includegraphics[width=0.77\textwidth]{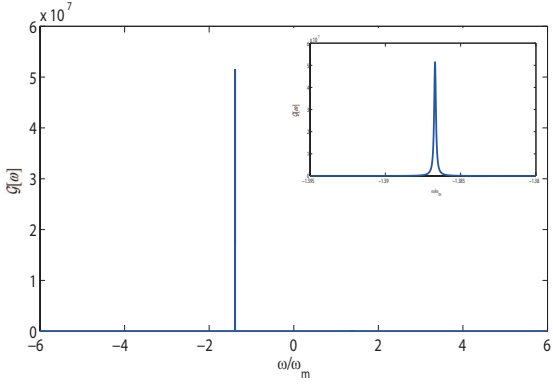}\\
  \caption{Power gain $\mathcal{G}[\omega]=|A(\omega)|^2$ as a function of the frequency $\omega$ where $\kappa_1=2.0\omega_m$, $\kappa_2=0$, $J=2.0\omega_m$, $G=1.0747\omega_m$ and $\Gamma_m=0.6\omega_m$. In the inset we plot the power gain $G$ as a function of $\omega$ around peak value.}\label{unresolvedg}
  \end{flushleft}
\end{figure}

\begin{figure}[htbp]
  \includegraphics[width=0.68\textwidth]{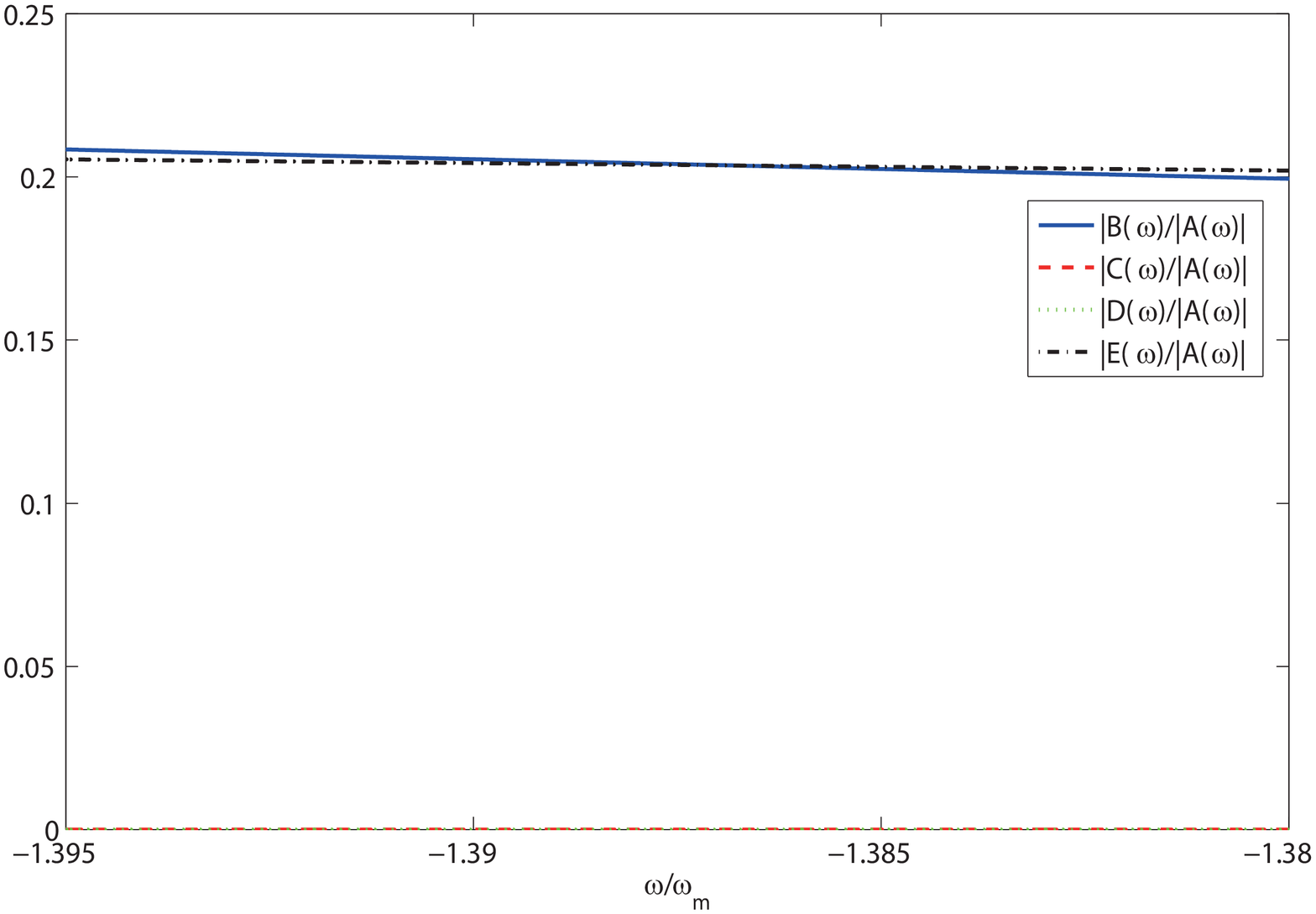}\\
  \caption{$|B(\omega|/|A(\omega)|$(blue solid line), $|C(\omega)|/|A(\omega)|$(red dashed line), $|D(\omega)|/|A(\omega)|$(green dotted line) and $|E(\omega)|/|A(\omega)|$(black dash dot line) as functions of $\omega$. The parameters are the same as in Fig. \ref{unresolvedg}. }\label{unresolved03}
\end{figure}

\begin{figure}[htbp]
  \includegraphics[width=0.68\textwidth]{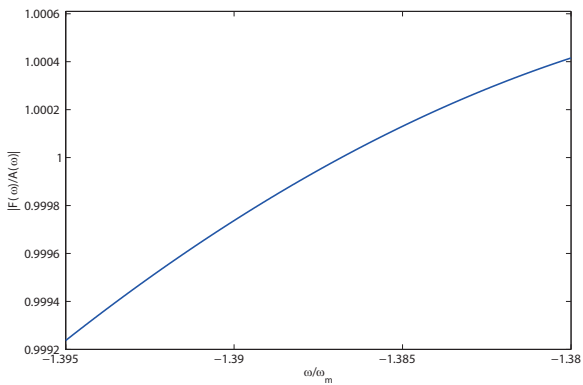}\\
  \caption{$|F(\omega)|/|A(\omega)|$ as a function of $\omega$. The parameters are the same as in Fig. \ref{unresolvedg}. }\label{unresolvedcontrast}
\end{figure}

 The power gain $\mathcal{G}[\omega]$ as a function of the frequency $\omega$ is shown in Fig. \ref{unresolvedg}. In Fig. \ref{unresolved03} we plot $|B(\omega)|/|A(\omega)|, |C(\omega)|/|A(\omega)|, |D(\omega)|/|A(\omega)|$ and $|E(\omega)|/|A(\omega)|$ as functions of the frequency $\omega$. An approximation $|B(\omega)|, |C(\omega)|, |D(\omega)|, |E(\omega)|\ll|A(\omega)|$ is appropriate. In Fig. \ref{unresolvedcontrast} we plot $|F(\omega)|/|A(\omega)|$ as a function of $\omega$. The number of added noise photons $N=(\overline{n}_{eff}+\frac{1}{2})(|F|^2/|A|^2)\approx \overline{n}_{eff}+\frac{1}{2}$. Since plenty of schemes have been proposed to cool the mechanical resonator to ground state in the unresolved sideband regime recently\cite{us1, us2}, our amplifier can reach the quantum limit and is phase insensitive.

Now we discuss the influence of non-zero $\kappa_2$. First we consider $\kappa_2=0.01\omega_m$ and other parameters the same as those in Fig. \ref{unresolvedg}. A quantum limited and phase-preserving amplifier can be obtained. The maximum gain power appears at $\omega_{p}=-1.39\omega_m$, the bandwidth is $2.6\times10^{-3}\omega_m$ and the gain-bandwidth product is $|\kappa|=0.51\omega_m$.

  In the situation where $\kappa_2<\omega_m$ but not $\kappa_2\ll\omega_m$, we give a set of parameters where $\kappa_2=0.5\omega_m$, $\kappa_1=3.0\omega_m$, $J=1.49\omega_m$, $\Gamma=0.2\omega_m$ and $G=0.5568\omega_m$. In this case the maximum gain power appears at $\omega_{p}=-1.08\omega_m$, the bandwidth is $6.0\times10^{-5}\omega_m$ and the gain-bandwidth product is $|\kappa|=0.21\omega_m$.
\subsection{tunable frequency window}
\begin{figure}
  \centering
  \includegraphics[width=0.7\textwidth]{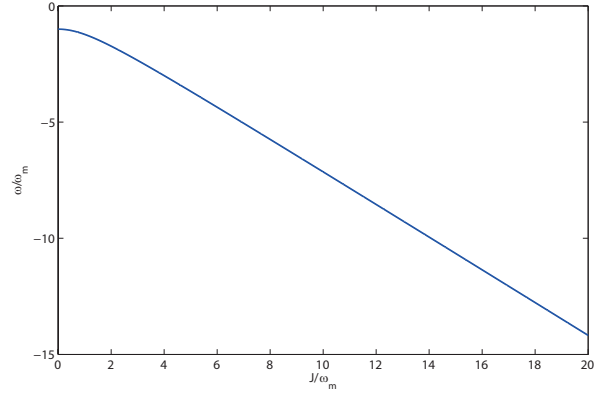}\\
  \caption{The center frequency as a function of $J$. Here $\kappa_1=\Gamma=0.1\omega_m$.}\label{tuam}
\end{figure}

The coupling coefficient $J$ of the compound cavity is tunable\cite{tu1, tu2, tu3, tu4, tu5, tu6, tu7, tu8}. The device features an amplifier with continuously tunable operating frequency. According to Eq. (\ref{omega1}) the center frequency appears at $\omega=-|\omega^{(1)}|$. In Fig. \ref{tuam} we plot the center frequency as a function of the two cavity coupling coefficient $J$.

\section{conclusion}
We propose a scheme to realize the phase-preserving amplification beyond the resolved sideband regime. Hence we can optimize the amplification in a large parameter range and achieve an enhancement of one magnitude in gain-bandwidth product. Also the frequency window of our proposed  amplifier can be tuned continuously by the adjustment of the cavity coupling.
\section*{ACKNOWLEDGMENTS}
We acknowledge W. J. Zou, F. C. Lei, M. Gao, Q. L. Jing and Y. H. Zhou for helpful discussion. We acknowledge the financial support in part by the 10000-Plan of Shandong province (Taishan Scholars), NSFC grant No.
11174177 and 11474182, and the National High-Tech Program of
China grant No. 2011AA010800 and 2011AA010803.


\begin{thebibliography}{99}
%\bibitem{nonlocality1}
%A. Einstein, B. Podolsky, N. Rosen, Phys. Rev. \textbf{47}, 777(1935).
\bibitem{co1}
M. Aspelmeyer, and T. J. Kippenberg, Rev. Mod. Phys. \textbf{86}, 1391 (2014).
\bibitem{co2}
P. Meystre, Ann. Phys. \textbf{525}, 215 (2013).
\bibitem{co3}
F. Marquardt, and S. M. Girvin, Physics \textbf{2}, 40 (2009).
\bibitem{co4}
T. J. Kippenberg, and K. J. Vahala, Science \textbf{321}, 1172 (2008).
\bibitem{cooling1}
J. D. Teufel, T. Donner, D. Li, J. W. Harlow, M. S. Allman, K. Cicak, A. J. Sirois,	J. D. Whittaker, K. W. Lehnert, and R. W. Simmonds, Nature \textbf{475}, 359 (2011).
\bibitem{cooling2}
J. Chan, T. P. Mayer Alegre, A. H. Safavi-Naeini, J. T. Hill, A. Krause, S. Gr$\ddot{o}$blacher, M. Aspelmeyer, and O. Painter, Nature \textbf{478}, 89 (2011).
\bibitem{cooling3}
F. Elste, S. M. Girvin, and A. A. Clerk, Phys. Rev. Lett. \textbf{102}, 207209(2009).
\bibitem{cooling4}
A. Xuereb, R. Schnabel, and K. Hammerer, Phys. Rev. Lett. \textbf{107}, 213604(2011).
\bibitem{cooling5}
X. Wang, S. Vinjanampathy, F. W. Strauch, and K. Jacobs, Phys. Rev. Lett. \textbf{107}, 177204(2011).
\bibitem{cooling6}
J. Restrepo, C. Ciuti, and I. Favero, Phys. Rev. Lett. \textbf{112}, 013601(2014).
\bibitem{sem1}
T. L. S. Collaboration, Nat. Phys. \textbf{7}, 962 (2011).
\bibitem{am1}
A. Nunnenkamp, V. Sudhir, A. K. Feofanov, A. Roulet, and T. J. Kippenberg, Phys. Rev. Lett. \textbf{113}, 023604 (2014).
\bibitem{am2}
A. Metelmann, and A. A. Clerk, Phys. Rev. Lett. \textbf{112}, 133904 (2014).
%\bibitem{sem1}
%A. Kronwald, F. Marquardt, and A. A. Clerk, Phys. Rev. A \textbf{88}, 063833 (2013).
%\bibitem{lv}
%X. Y. L$\ddot{u}$, J. Q. Liao, L. Tian and F. Nori, Phys. Rev. A \textbf{91}, 013834 (2015).
\bibitem{sem2}
E. Gavartin, P. Verlot, T. J. Kippenberg, Nature nanotechnology, \textbf{7}, 509 (2012).
\bibitem{hs1}
M. Aspelmeyer, P. Meystre, and K. Schwab, Phys. Today \textbf{65}, 29 (2012).
\bibitem{hs2}
G. De Chiara, M. Paternostro, and G. M. Palma, Phys. Rev. A, \textbf{83}, 052324 (2011).
\bibitem{hs3}
\bibitem{mp}
C. M. Caves, Phys. Rev. D \textbf{26}, 1817 (1982).
\bibitem{qf}
H. A. Haus, \emph{Electromagnetic Noise and Quantum Optical Measurements}, (Springer, Berlin, 2000), p. 267-277.
\bibitem{sc1}
M. A. Castellanos-Beltran, K. D. Irwin, G. C. Hilton, L. R. Vale, and K. W. Lehnert, Nature Phys. \textbf{4}, 929 (2008).
\bibitem{sc2}
N. Bergeal,	F. Schackert, M. Metcalfe, R. Vijay, V. E. Manucharyan,	L. Frunzio,	D. E. Prober, R. J. Schoelkopf,	S. M. Girvin and M. H. Devoret, Nature \textbf{465}, 64 (2010).
\bibitem{us1}
Y. C. Liu, Y. F. Xiao, X. Luan, Q. Gong, and Chee Wei Wong, Phys. Rev. A \textbf{91}, 033818 (2015).
\bibitem{us2}
Y. Guo, K. Li, W. Nie, and Y. Li, Phys. Rev. A \textbf{90}, 053841 (2014).
\bibitem{rhc}
E. X. DeJesus, and C. Kaufman, Phys. Rev. A \textbf{35}, 5288 (1987).
\bibitem{ps}
A. A. Clerk, M. H. Devoret, S. M. Girvin, F. Marquardt, and
R. J. Schoelkopf, Rev. Mod. Phys. \textbf{82}, 1155 (2010).
\bibitem{tu1}
I. S. Grudinin, H. Lee, O. Painter, and K. J. Vahala, Phys. Rev. Lett. \textbf{104}, 083901(2010).
\bibitem{tu2}
Q. Xu, S. Sandhu, M. L. Povinelli, J. Shakya, S. Fan, and
M. Lipson, Phy. Rev. Lett. \textbf{96}, 123901(2006).
\bibitem{tu3}
J. Cho, D. G. Angelakis, and S. Bose, Phys.Rev.A \textbf{78}, 022323(2008).
\bibitem{tu4}
I. S. Grudinin and K. J. Vahala, Opt. Express \textbf{17}, 14088(2009).
\bibitem{tu5}
Y.-F. Xiao, M. Li, Y.-C. Liu, Y. Li, X. Sun, and Q. Gong,
Phys.Rev.A \textbf{82}, 065804(2010).
\bibitem{tu6}
C. Zheng, X. Jiang, S. Hua, L. Chang, G. Li, H. Fan, and
M. Xiao, Opt. Express \textbf{20}, 18319(2012).
\bibitem{tu7}
Y. Sato, Y. Tanaka, J. Upham, Y. Takahashi, T. Asano, and
S. Noda, Nat. Photon. \textbf{6}, 56(2012).
\bibitem{tu8}
B. Peng, $\c{S}$. K. $\ddot{O}$zdemir, F. Lei, F. Monifi, M. Gianfreda,
G. L. Long, S. Fan, F. Nori, C. M. Bender, and L. Yang, Nat.
Phys. \textbf{10}, 394(2014).
\end{thebibliography}
\end{document}